\documentclass[12pt]{article}

\usepackage{amsmath,amssymb,graphicx,mathrsfs,slashed,setspace}

\usepackage{float}
\usepackage{latexsym}
\usepackage{bm}
\usepackage{blindtext}
\usepackage{hyperref}
\hypersetup{
	colorlinks=true,
	linkcolor=blue,
	filecolor=black,
	urlcolor=black,
	citecolor=blue,
}
\usepackage{comment}

\newcommand{\hide}[1]{}

\newcommand{\be}{\begin{equation}}
\newcommand{\ee}{\end{equation}}
\newcommand{\bea}{\begin{eqnarray}}
\newcommand{\eea}{\end{eqnarray}}

\def\({\left(}
\def\){\right)}

\renewcommand{\baselinestretch}{1.5}
\begin{document}
%%%%%%%%
\title{\vspace{-1.8in}
%\vspace{3mm}
%\vspace{0.3cm}
{Formation of Frozen Stars from \\ collapsing matter by tunneling }}
\author{\large Ram Brustein${}^{(1)}$,  A.J.M. Medved${}^{(2,3)}$, Tamar Simhon${}^{(1)}$
	\\
	\vspace{-.5in} \hspace{-1.5in} \vbox{
		\begin{flushleft}
			$^{\textrm{\normalsize
					(1)\ Department of Physics, Ben-Gurion University,
					Beer-Sheva 84105, Israel}}$
			$^{\textrm{\normalsize (2)\ Department of Physics \& Electronics, Rhodes University,
					Grahamstown 6140, South Africa}}$
			$^{\textrm{\normalsize (3)\ National Institute for Theoretical Physics (NITheP), Western Cape 7602,
					South Africa}}$
			\\ \small \hspace{0.57in}
  ramyb@bgu.ac.il,\  j.medved@ru.ac.za,\  simhot@post.bgu.ac.il
\end{flushleft}
}}
\date{}
\maketitle
%%%

%%%%%%%%%%%%%%
%\renewcommand{\baselinestretch}{1.15}
%%%%%%%%%%%%%%%%%
%%%%%%%%%%%%%%%%%%

\begin{abstract}
%\abstract
The frozen star is a type of black hole mimicker: An ultracompact object whose exterior geometry resembles that of a general-relativistic black hole but differs in its  matter composition and in the regularity of its interior geometry.  It is sourced by a spherically symmetric collection of open-string flux tubes, which posses an extremely anisotropic energy-momentum-stress tensor with maximally negative radial pressure. The frozen star  represents an effective classical description of the highly quantum, closed-string polymer model. A key challenge for any model of a black hole mimicker  is to explain how such objects can form from a collapsing body of matter. We started to address this important problem in \cite{U4Euclidean} by adapting the Euclidean-action method of Gibbons and Hawking to show that the transition into a frozen star is likely. Here, we improve on our previous results by showing that the transition probability for a collapsing shell of matter to tunnel quantum mechanically into a frozen star is unity, up to negligible  corrections. Our conclusion is that such a transition is therefore  inevitable.
\end{abstract}
\maketitle

\newpage
\renewcommand{\baselinestretch}{1.5}
\setstretch{1.5}
\section{Introduction}

The paradoxical issues that are associated with black hole (BH) singularities and  horizons \cite{FW} have prompted many  to propose models of ultracompact objects that are devoid of  singularities and trapped surfaces but, at the same time, would look just like standard (general relativistic)  BHs when viewed from the exterior \cite{carded,BHmimickers,Carballo-Rubio:2025fnc}. An example of such a BH mimicker model is the frozen star model, as first described in \cite{bookdill,BHfollies} and further developed in \cite{popstar,trajectory,fluctuations,U4Euclidean}. More recent progress includes the incorporation of rotation  to obtain a Kerr BH mimicker \cite{notstevekerr}, the inclusion of a Lagrangian for the matter source \cite{StringFluid}  and the combination thereof \cite{StringPlusRot}. Also see \cite{frosty2}.

A frozen star seems an unlikely candidate for mimicking an astrophysical BH. It is sourced by a fluid of maximally negative radial pressure such that the radial component of the null-energy condition (RNEC) is saturated throughout. Furthermore, the fluid  is highly anisotropic: Its transverse pressure components are vanishing. While the star does not possess a formal horizon, its outermost surface along with every radial slice in its interior has an exponentially large redshift.  This special blend of properties, as exotic as they might sound, are just what is needed for an ultracompact  object to mimic two essential features of a standard
BH. First,  the  interior geometry and matter within a frozen star  are ultrastable against perturbations \cite{bookdill,popstar,fluctuations,frosty2}, meaning that, just like for a standard BH,  both sides of the linearized Einstein equations are identically vanishing and, second,   the star's entropy  has been shown \cite{U4Euclidean} to satisfy  the
BH area--entropy law of Bekenstein and Hawking \cite{Bek,Haw}.

In fact, any ultracompact object with an exponentially large redshift is endowed with a temperature
that is perturbatively close to the Hawking value \cite{adiabatic,Samir,samir2} and, consequentially,
by virtue of the first law of thermodynamics, with an entropy that is perturbatively close to  the Bekenstein--Hawking prescription. One can also derive the same result  by using the
Euclidean path-integral method of Gibbons and Hawking \cite{GH}, which can be generalized to include objects whose
interior composition differs from that of conventional matter and/or those having  unusually low
temperatures.~\footnote{These loopholes are clarified in the original discussion \cite{GH}, just above and below their Eq.~(3.16).} The key
to this  Gibbons--Hawking result is that, as long as the interior contribution to the entropy is
parametrically smaller than that of the area law, only the asymptotic boundary term in the exterior
spacetime contributes to the calculation at leading order. We were able to show in
\cite{U4Euclidean} for the frozen star model that the integration of the Euclidean action in the bulk of the star's interior
 reproduces the very same  area--entropy law but, thanks to the cancelling effects of an
outer transitional  layer, the interior still makes no net contribution to the standard calculation
at the asymptotic boundary. In this way, we have argued that the frozen star is, to the best of our
knowledge, the only BH mimicker whose  interior composition is self-consistent with the generalized
Gibbons--Hawking calculation.

Since the appearance of \cite{U4Euclidean}, we have identified the matter which sources the frozen
star geometry \cite{StringFluid}.  This matter is described by the  Lagrangian for a
particular string fluid \cite{GHY,Yi}
that results from the decay of an unstable $D$-brane or a brane--antibrane system at the end of
open-string tachyon condensation, as originally described by Sen \cite{Sen,Sen2,Sen3,Sen4,Sen5}.
When Sen's Lagrangian is recast into a Born--Infeld form, as in \cite{GHY}, and then
coupled to gravity, the static and spherically symmetric solution reveals a picture of rigid,
radially directed  tubes carrying lines of electric flux that extend from a point-like source in the  core
of the star to a spherical distribution of equal and opposite charge on its exterior
surface.~\footnote{Except for the charges  and flux, a similar picture of the frozen star interior
was discussed in \cite{trajectory} and described, following \cite{hedge1,hedge2}, as a ``hedgehog
compactification''.} In this sense, a frozen star is a gravitationally compactified BIon
\cite{Gibbons:1997xz,GibbonsRev}; that is, a BIon whose flux lines do not extend beyond its outer
surface because of the gravitational back reaction. It is interesting to recall that the underlying
premise of the frozen star is to provide an effective  classical description of the strongly quantum
polymer model, which  describes an ultracompact object whose interior consists of a highly excited
fluid of closed, rather than open, strings \cite{inny,strungout,emerge}.

The main purpose of the current paper is to improve on a previous analysis
which  showed that it
is likely for a collapsing body of matter to transition into a frozen star \cite{U4Euclidean}.
To this end, we will evaluate the
Euclidean partition function of the frozen star, including the
contribution from the Born--Infeld matter Lagrangian and
the quadratic corrections to the zeroth-order result. On this basis, we
are able to calculate the probability
for a collapsing shell of matter to transition into a  frozen star
via a quantum-tunneling process. This probability is shown to be unity, up to
negligibly small corrections, from
which it can be concluded   that
such a transition is  inevitable! This realizes in detail a proposal by Mathur, which was
made in the context of fuzzballs \cite{mathur}.

The rest of the paper proceeds as follows: The geometry of the frozen star model is briefly reviewed in Section~2. In Section~3,  we recall the results of \cite{StringFluid}, where it is shown that the solutions of the equations of motion for the Born--Infeld Lagrangian of \cite{GHY} coupled to Einstein--Hilbert gravity reproduce the frozen star geometry. Additionally, an outer-surface transitional layer is reconsidered from the Born--Infeld perspective; in particular, inevitably large values of  transverse pressure in the layer \cite{Wilty} are related to the outer distribution of the electric charge. Next, a novel determination of the radial profile of the (inverse) temperature is covered in Section~4. This outcome further justifies our choice of a radially dependent Euclidean compactification scale, which was essential to the reproduction of the area--entropy law in \cite{U4Euclidean} and is equally important to the current analysis. In Section~5, which is the most important section, we establish the main result. We calculate the transition probability from a collapsing matter shell into a frozen star and show that it is unity. The final section contains a brief overview.

%%%%%%%%%%%%%%%%%%%%%%%%%%%%%%%%%%%%%%%%%%%%%%%%%%%%%%%%%%%%%%%%%%%

\section{The frozen star model}\label{FS}

A brief review of the frozen star model is in order.  Spherical symmetry and  staticity are assumed,~\footnote{The recent extension to rotating, stationary solutions \cite{notstevekerr,StringPlusRot} will
not be considered.} although it is expected that the first implies the second via a Birkhoff-like theorem due to the star's ultrastability.  For the bulk of the compact object, the frozen star geometry  is described by  the simplest spherically symmetric and static metric that  results in the saturation of the RNEC and the vanishing of the transverse pressure throughout the interior. The last two conditions are motivated by  similar properties of the antecedent  polymer model as discussed at length in \cite{bookdill,BHfollies}. The corresponding line element is given by
\be
ds^2\;=\; -\varepsilon^2 dt^2+ \frac{1}{\varepsilon^2} dr^2+ r^2d\Omega^2\;.
\label{FSmetric}
\ee
Here, $\;\varepsilon^2\ll 1\;$ is a dimensionless, constant parameter that should be
regarded as exponentially small. It then follows that  the geometry is almost null throughout the interior.  It is straightforward to show that, if $M$ is the star's mass and $R$ is its radius, then $\;R=2MG(1+\varepsilon^2)
+\mathcal{O}(\varepsilon^4)\;$.

The resulting Einstein tensor is diagonal, whose components correspond to the following matter densities via the Einstein equations,
\bea
\label{polyrho}
8\pi G\;   \rho &=& \frac{1-(rf)'}{r^2} \;=\;  \frac{1-\varepsilon^2}{r^2}\;, \\
\label{polypr}
8\pi G \;  p_r &=& -\frac{1-(rf)'}{r^2} \;=\; -\frac{1-\varepsilon^2}{r^2}\;, \\
\label{polypt}
8\pi G \;  p_\perp &=& \frac{(rf)''}{2r}\;=\; 0\;,
\eea
where $\rho$ is the energy density, $p_i$ is a component of pressure and a prime denotes a radial derivative. The RNEC saturation condition is $\;\rho +p_r=0\;$.

The stress-tensor conservation equation reduces from its general static and spherically
symmetric form,
$\; p'_r+ \frac{1}{2}(\ln{f})'\left(\rho+p_r\right)+\frac{2}{r}\left(p_r- p_{\perp}\right)=0\;$, to
\be
p_{\perp}\; = \;\frac{1}{2 r} \partial_r(r^2 p_r)\;.
\label{conserv2}
\ee

There are two special regions, the central core  and outermost layer of the star, that deviate from the forms as presented above. First, a very small region close to the center has to  be regularized so that the relevant densities and curvature invariants remain finite \cite{trajectory}.  By taking the radius $2\eta$ of the regularized sphere to be sufficiently small, $\;\eta\ll R\;$, it is confirmed that integrated   quantities like the mass only deviate from their bulk values by corrections of relative order $\;\frac{\eta}{R}\;$ or higher. For this reason, the regularized core is only currently relevant to the value of the temperature near $\;r\sim\eta\;$, which is discussed in Section~4. Further note that the RNEC saturation condition  can be  maintained in the regularized region.

The transitional layer  at the outermost surface  of the star is necessary so that the internal geometry can be smoothly connected to the external Schwarzschild solution. Like the regularized sphere, we regard the width $2\lambda$ of this  layer to be narrow, $\;\lambda \ll R\;$  (although  larger than  $\;\varepsilon^2 R\;$) and choose to maintain the condition $\;\rho+p_r=0\;$ throughout the layer.  Unlike the central core, though, the price of this condition in the outer layer is that the transverse pressure grows quite large, $\;p_{\perp} \sim \frac{R}{\lambda}\rho\gg \rho\;$, as follows in part  from Eq.~(\ref{conserv2}). Consequently, some integrated quantities can deviate greatly from their bulk-only values, although the total mass is not one of them.

The RNEC saturation condition is essential to some of the important features of the frozen star solution. These include its  ultrastability, meaning that all radial perturbations of the background geometry vanish identically or decay instantly \cite{bookdill,popstar}, and the same can be deduced from the results of \cite{fluctuations,frosty2} for the angular perturbations.  Additionally, its ability to evade both the singularity theorems \cite{PenHawk1,PenHawk2} and the compactness-of-matter bounds \cite{Buchdahl,chand1,chand2,bondi}.  In the polymer model, the analog of the RNEC saturation is the condition of maximal entropy \cite{strungout}, which leads to similar features.

%%%%%%%%%%%%%%%%%%%%%%%%%%%%%%%%%%%%%%%%%%%%%%%%%%%%%%%%%%%%%%%%%%%

\section{Frozen stars as gravitationally back-reacted BIons}

Here, we recall some of the main results from \cite{StringFluid}, focusing on what is needed to understand  the subsequent sections. The new input from \cite{StringFluid} is  the inclusion of a matter Lagrangian with a Born--Infeld form, which can be motivated from a string-theoretical perspective and is based on a framework that was first put forth by Gibbons, Hori and Yi \cite{GHY} (also see \cite{Yi}).

The resulting picture is a frozen star interior that can be viewed as a collection of rigid tubes of electric flux that extend from a point-like charge at the center of the star up to a spherical charge distribution of equal and opposite charge on the outer surface. As discussed in \cite{StringFluid}, the attractive electric force between the  two oppositely charged distributions  is exactly cancelled  by a repulsive ``Lagrange-multiplier'' force that arises due to the constraint of fixed mass. This also cancels the dipole moment. As the star is net neutral and the distribution is spherically symmetric, the configuration is devoid of any higher-order multipole moments, so that the exterior is assured of being in its  standard Schwarzschild vacuum state.

Up to  surface terms (in particular, the Lagrange-multiplier term discussed above), the total --- Einstein--Hilbert (EH) plus Born--Infeld (BI)  --- action is now
\bea
&& S_{EH+BI}=
\int d^4 x\left\{{\cal L}_{EH} + {\cal L}_{BI} \right\} \label{GBI} \\ && =
\int d^4 x \left\{\frac{1}{16 \pi G} \sqrt{-g}R^{a}_{\;\;a} +\frac{1}{2\pi \alpha'}\sqrt{-\frac{1}{2}{\cal K}^{ab}{\cal K}_{ab}}+\lambda
\varepsilon^{abcd}{\cal K}_{ab}{\cal K}_{cd}
+  \sqrt{-g} J_a A^a \right\} \;, \nonumber
\eea
where $\alpha'$  is the inverse of the fundamental string tension and ${\cal K}_{ab}$ is an effective field-strength tensor (actually, a tensor density), which
is related to the fundamental field-strength tensor ${\cal F}_{ab}$
by way of a canonical transformation. The second term on the right
is the main portion of the Born--Infeld Lagrangian, while the third  one is another Lagrange-multiplier term  which is needed to enforce the constraint
$\;{\cal K} \wedge {\cal K} =0\;$ and  the fourth one is the source term which includes a 4-current $J_a$ and a gauge field $A^a$.
Importantly, $A^a$ is the gauge field
for ${\cal F}_{ab}$ but  {\em not} ${\cal K}_{ab}$, whose gauge
field we rather denote by $\widetilde{A}^a$. The two gauge fields
are assumed to be independent.

The only sources that we consider here are the point-like charge $q_{core}$ at the origin and the equal but oppositely
charged spherical distribution at the outer surface. With this choice of sources, the Born--Infeld portion of the Lagrangian in the bulk and the  energy density  both reduce to the same
simple expression ($i,j,\dots$ denotes a spatial index),
\begin{equation}
	\frac{1}{\sqrt{-g}}{\cal L}_{BI}\;=\; \rho \;=\; E_iD^i\;,
	\label{BILan}
\end{equation}
where $\;E_i= \delta_i^{\;\;r}\partial_r A^0\;$
and $\;D_i = \delta_i^{\;\;r}\partial_r \widetilde{A}^0\;$ are, respectively,  the electric and  displacement
fields for ${\cal F}_{ab}$. Alternatively, $D_i$ is the electric field
for ${\cal K}_{ab}$. As anticipated, one also finds that $\;p_r=-\rho\;$ and $\;p_{\perp}=0\;$.

To ensure that the diagonal stress-tensor elements agree with their counterparts from the Einstein tensor,  while
maintaining the standard relation between the electric and displacement field for a Born--Infeld theory, one obtains
that
\begin{equation}
	D^r\;=\;\frac{q_{core}}{4\pi r^2}\;
	\label{DR}
\end{equation}
and
\begin{equation}
	E_r\;=\;\frac{1}{2\pi \alpha'}\;,
	\label{ER}
\end{equation}
where $\;q_{core}=\pi \frac{\alpha'}{G}\;$ is independent of the size of the  star.

In this framework, it is the displacement field that satisfies the Gauss'-law constraint, as the variation of the total Lagrangian in Eq.~(\ref{GBI})  by $A^{a}$ leads directly to the desired expression $\;\nabla_i D^i= J_0=\rho_e\;$, where $\rho_e$ is the volume charge density. A more extensive discussion on the Born--Infeld  field equations can be found in either \cite{StringFluid} or \cite{GHY}.

%%%%%%%%%%%%%%%%%%%%%%%%%%%%%%%%%%%%%%%%%%%%%%%%%

\subsection{Transverse pressure as a source}\label{ch4}

We  now want to understand the unusually large transverse pressure in the outer transitional layer, as discussed at the end of Section \ref{FS}, in terms of the flux tubes which source the frozen star solution. To this end, we assume  that the charge at the surface is distributed uniformly throughout this narrow layer of width $2\lambda$. Hence, at leading order in the perturbative parameters $\lambda/R$ and $\varepsilon^2$,
\be
J_0\;=\;\rho_e \;= \;-\frac{q_{core}}{8\pi R^2 \lambda}\;.
\label{sourcey}
\ee

Let  $\;{\cal L}_{source}=\sqrt{-g} J_aA^a=\sqrt{-g}J_0A^0\;$ denote the source term of the Lagrangian. The contribution to the stress tensor from this source term is then
\begin{equation}
	T_{\mu\nu}^{source}\;=\;\frac{2}{\sqrt{-g}}\frac{\delta {\cal L}_{source}}{\delta g^{\mu\nu}}\;=-g_{\mu\nu}J_0 A^0+ 2 A_{\mu} J_\nu \;.
\label{Tmunu}
\end{equation}

Restricting the  previous equation to spatial components,
we then have
\begin{equation}
	T^{i\  source}_{\;\;j}\;=\;-\delta^{i}_{\;\;j} J_0 A^0\;.
	\label{nolabel}
\end{equation}
This expression represents  the sole contribution to
the  angular components  of  the total stress tensor because there can be no
such contribution from the main part of the Born--Infeld Lagrangian without breaking
one or more of spherical symmetry and staticity \cite{Letel,GHY}.
Hence,
\be
p_{\perp}\;=\;  T^{\theta}_{\;\;\theta}
\;=\; T^{\phi}_{\;\;\phi}\;=\; -J_0A^0 \;.
\label{rel1}
\ee
Equation~(\ref{sourcey}) for the charge density requires $J_0$ to scale with $1/\lambda$, which explains why $p_{\perp}$ has to scale with $1/\lambda$
in the transitional  layer from the Born--Infeld perspective.~\footnote{That $A_0$ itself has no $\lambda$ dependence follows from the simple form of the electric field in Eq.~(\ref{ER}).}
	
%%%%%%%%%%%%%%%%%%%%%%%%%%%%%%%%%%%%%%%%%%%%%%

\section{Temperature of the frozen star}

In this section, we calculate the temperature of a frozen star using the heat equation. In \cite{U4Euclidean}, it was argued that the inverse temperature should scale linearly with the radial coordinate $r$. We now show that this very same scaling arises because the frozen star is static and therefore in thermal equilibrium.

In a static and  spherically symmetric gravitational background, the heat equation takes the form
\begin{equation}
	\nabla^2 {\widetilde T}(r)\;=\frac{1}{\sqrt{-g}}\partial_r\left(\sqrt{-g}g^{rr}\partial_r {\widetilde T}(r)\right)=\;0\;,
\label{heat}
\end{equation}
where ${\widetilde T}(r)$ is the local  temperature; that is, it includes the Tolman blueshift factor of $\frac{1}{\sqrt{|g_{tt}|}}$. See \cite{santyclaws} for a modern discussions on the Tolman temperature and its gradients.
In our case,  $\;\sqrt{-g}= r^2\;$ and  $\;g^{rr}=|g_{tt}|=\varepsilon^2$\;.

The boundary conditions supplementing the heat equation are fixed at infinity and at the outer surface of the star (technically, the inner surface of the transitional layer). The temperature at the outer surface
has to be perturbatively close to the Hawking value because
of arguments presented in \cite{U4Euclidean} that follow along the lines of
a more general discussion in \cite{adiabatic} (also see \cite{Samir}).
With this in mind, we solve the heat equation with the  boundary conditions
\bea
{\widetilde T}(r\rightarrow \infty)&=&\frac{1}{4\pi R}\;,
\label{bcinf} \\
{\widetilde T}(r = R-\lambda)&=&\frac{1}{4\pi (R-\lambda) \sqrt{\varepsilon^2}}\;.
\label{bcR}
\eea

The general solution of Eq.~(\ref{heat}) is the following:
\be
{\widetilde T}(r)\;= \;\;C_1~ \frac{1}{r} +C_2\;,
\ee
where $C_1$ and $C_2$ are constants to be determined by the conditions~(\ref{bcinf},\ref{bcR}). Our particular  solution is given by
\begin{equation}
	{\widetilde T}(r)\;=\;\frac{1-\varepsilon}{4\pi \varepsilon r}+\frac{1}{4\pi R}+\mathcal{O}(\lambda)\;,
\end{equation}
where corrections of order $\varepsilon^2$ have been left as implied.
Multiplying the local temperature by the Tolman factor  $\varepsilon$ so as to
obtain the externally measured temperature, we end up with
\begin{equation}
T(r)\;=\;\frac{1}{4\pi  r}+\mathcal{O}(\varepsilon)\;.
\end{equation}
Note that, near $\;r=0\;$, one has to use the regularized core of the frozen star, leading  to a regularized, large-but-finite temperature.

This result can be compared to the stationary heat conduction result
that one obtains  when  the two ends of a rod are kept at different temperatures by thermostatic devices. In our case,  heat naturally  flows from the hotter end, $\;T_{bulk}=\frac{1}{4\pi r}$ in the bulk  of the star, to the cooler one, $\;T_{\infty}=\frac{1}{4\pi R}\;$ at the surface of the star.

The inverse temperature $\;\beta(r)=\frac{1}{T(r)}\;$,  for $r$ of order $R$ or smaller, is then given by
\be
\beta \;=\;4\pi r\;,
\label{betaE}
\ee
up to corrections of order $\varepsilon$.
As already mentioned, we arrived at the same result using arguments about the Euclidean time-periodicity scale in \cite{U4Euclidean}.

%%%%%%%%%%%%%%%%%%%%%%%%%%%%%%%%%%%%%%%%%%%%%%%%%
\section{Dynamical formation of frozen stars from a collapsing shell of matter}

Our main objective is to calculate the probability of a quantum transition from
a collapsing shell of matter to a frozen star of the same mass.

One might wonder how a collapsing system of standard-model matter could evolve, by classical gravitational dynamics,  into  an ultracompact object whose composition  is that of exotic stringy matter and whose geometry  deviates from the standard Schwarzschild description on horizon-sized  length scales. The short answer is that it couldn't. In \cite{U4Euclidean}, we proposed, following an idea first introduced by Mathur in the context of fuzzballs \cite{mathur},  that the correct description of this evolution must include a quantum-induced phase transition or, equivalently, a quantum tunneling event. We then argued that, from this perspective, it is natural to regard the Euclidean version of the outer transitional layer as a quantum-gravitational instanton which is mediating the transition from the empty interior of an infalling shell of  conventional (standard model) matter to a same-sized sphere of exotic  frozen star matter. For some relevant discussions in the literature on quantum-gravitational instantons, see \cite{coleman,HH,GGS,BC,Brown:2007sd}.

The original idea in \cite{U4Euclidean} was that, once the collapsing matter shell reaches  the outermost edge of the transitional layer, Euclidean evolution is triggered and continues  until the shell reaches the innermost edge of the layer. At this point, Lorentzian evolution resumes and the bulk of the frozen  star forms dynamically by some yet-to-be-determined process.

To understand the triggering mechanism, consider that the inside of the shell is decaying from the false vacuum
of Minkowski space to the true vacuum of a frozen star, as dictated by the large discrepancy in their respective values of entropy:
zero versus $\;S_{FS}=S_{BH}\;$.  The instanton then acts like a
domain wall and  induces a bounce from a bubble of  nothing
to the  entropically preferred  final state. The reason that the transition is triggered when the shell is just outside of its Schwarzschild radius, as determined
by the spatial location of
the narrow transitional layer, is
because this is when its wavefunction has the energy to jump over the potential barrier into a microstate of the frozen star. The key being that  local temperatures
grow to be exponentially large in the limit of the would-be horizon \cite{bw}.

In the current calculation, we include the bulk of the frozen star in the instanton as a means of summing over all possible microstates, which was done by hand in the original account. The added bonus of this  perspective is that the formation of the bulk of the star becomes part of the Euclidean tunneling  picture and is not left to some unknown Lorentzian stage of the evolution.

\begin{figure}[h]
\vspace{-.2in}
\hspace{1in}
	\includegraphics[height=8cm]{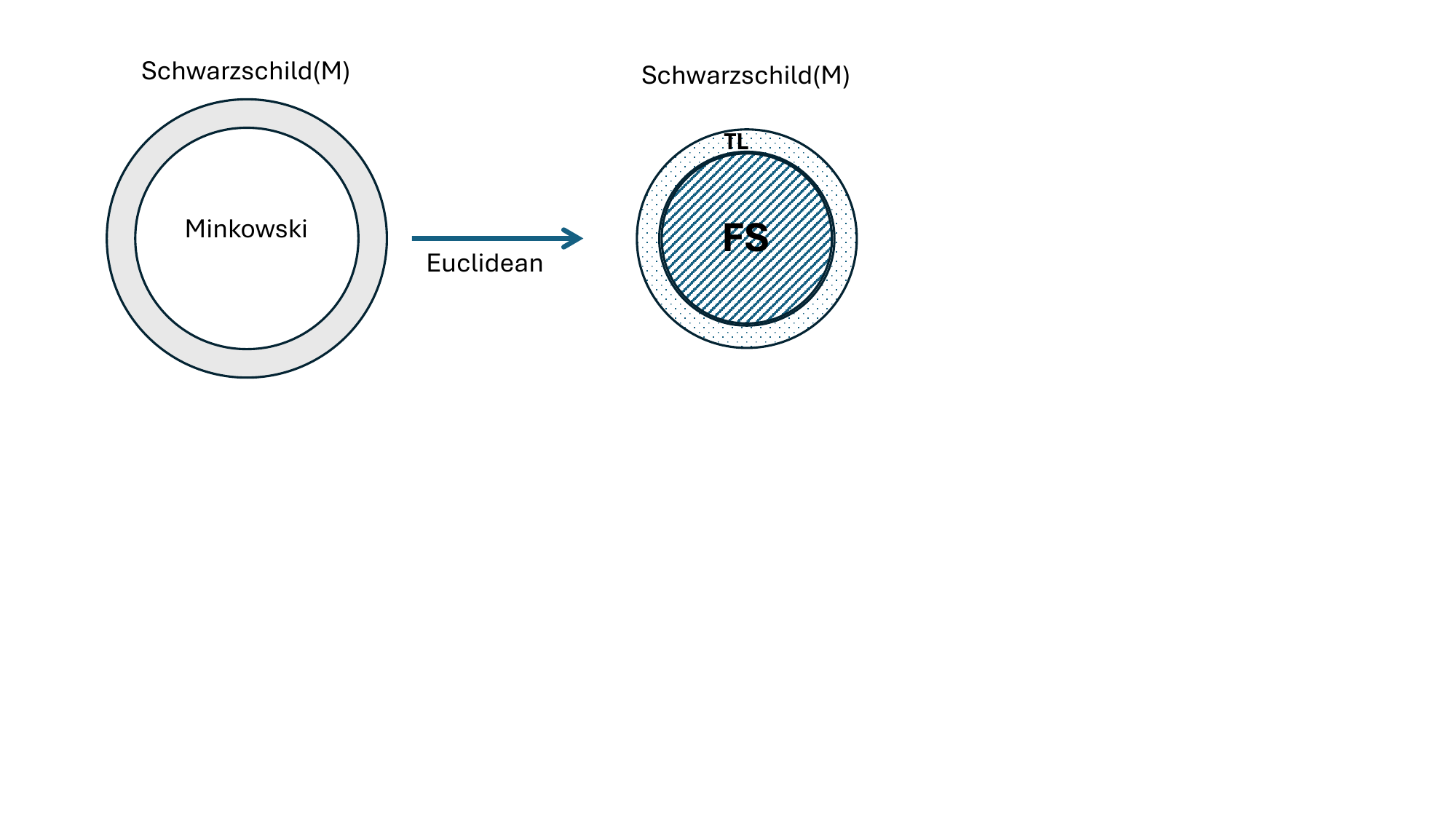}
\vspace{-2in}
	\caption{The collapsing shell tunneling to the true vacuum of the
	frozen star (FS), rather than continuing with its classical evolution.
	The transitional layer (TL) is not drawn to scale.}
\label{tunneling}
\end{figure}

To sum up, as the shell of matter is  approaching its Schwarzschild radius, reaching a distance of $\lambda$ away, the huge entropy of the frozen star induces a tunneling event.
This is depicted in Figure~\ref{tunneling}.

Let us now introduce the partition function for a Euclidean instanton \cite{camb,FV1},
\begin{equation}
	{\cal Z}\;=\;\sum_i N_i\;\text{det}^{-\frac{1}{2}}\left( I''(x^0_{i}) \right)e^{-I(x^{0}_{i})}\;.\label{steep}
\end{equation}
Here,  $I$ is a Euclidean action, $x^{0}_{i}$ indicates the $i$-th stable stationary point of the action, $N_i$ is a normalization factor and a prime in this context represents a variation of the action with respect to one of its constituent fields. Importantly, one can regard the partition function as determining the probability of the transition that the instanton mediates,
 $\;\Gamma={\cal Z}\;$.

Just like in the case of a false vacuum (FV) decaying into a true vacuum (TV), we can construct such a partition function to describe the tunneling transition of a spherical shell of mass $M$, with a Minkowski interior, into the interior of a frozen star (bulk plus transitional layer) of the same mass.
Suitably refining Eq.~(\ref{steep}), we then have \cite{winey}
\begin{equation}
	{\cal Z}\;=\;{\cal A} \left|\frac{\text{det}'\;I''(\phi_{TV})}{\text{det}'\; I''(\phi_{FV})}\right|^{-\frac{1}{2}} e^{-(I_{TV}-I_{FV})}\;, \label{gammy}
\end{equation}
where ${\cal A}$ is a  constant which is related to the zero modes of the vacuua
and the primes on the determinants indicate that the zero modes should be excluded.~\footnote{Any zero mode results in the vanishing of the determinant and thus is accounted for  separately via the normalization
constant.} Also, $I_{TV}$ and $I_{FV}$  are the actions evaluated
on the respective background solutions.

As discussed in the Introduction, we have found previously that the probability for transition from the collapsing matter system to a frozen star is a number of
order unity. We will now proceed to calculate this transition probability more precisely.

\subsection{Einstein-Hilbert action}\label{EHaction}

Here, we briefly review the relevant results of \cite{U4Euclidean}, which applied strictly to the Einstein--Hilbert  part of the Euclidean action of the frozen star,
so as to make our discussion self-contained.

Let us start with the bulk of the frozen star interior ({\em i.e.}, up
to but not including the transitional layer). The Euclidean action of the bulk  does not receive any contribution from  surface terms because the area of a radial slice  vanishes at $\;r=0\;$   and the outermost surface  of the bulk is not a true boundary of spacetime. The remaining volume integral is
\bea
I_{EH, Bulk} &=& -\frac{1}{16\pi G}\int_{Bulk} d^4x\;\sqrt{g} \; R^{\mu}_{\;\;\mu}\;,
\eea
and, using  Eq.~(\ref{betaE}) for  the compactification scale of the Euclidean time direction, one can rewrite this as~\footnote{Corrections of order $\varepsilon^2$ and $\eta$ are implied in equations throughout Section~5.}
\be
I_{EH, Bulk} \;=\;
 -\frac{\pi}{G} \int_0^{R-\lambda} dr \; r^2 \;r R^{\mu}_{\;\;\mu}\;.
\ee

In the bulk of the frozen star,  $\;R^{\mu}_{\;\;\mu} = -8\pi G T^{\mu}_{\;\;\mu}=
-8\pi G\left(-\rho+p_r+2p_{\perp}\right)=16\pi G \rho=\frac{2}{r^2}\;$.  So that
\be
I_{EH, Bulk} \;=\;
 - 2\frac{\pi}{G} \int_0^{R-\lambda} dr  \; r  \;=\;  -\frac{\pi R^2}{G}
 + \mathcal{O}(\lambda) \;.
\label{lights,camera}
\ee

Moving on to the transitional layer, we recall that this is a thin shell of width $\;2\lambda \ll R\;$ whose  metric  (and its first two derivatives) smoothly connects  the interior bulk  and exterior Schwarzschild geometries.  Here, again, we only need to evaluate the volume term in the action because prospective surface terms are supported only on the true boundaries at $\;r=0\;$ and $\;r\to\infty\;$. For this layer, $\;p=-\rho\;$ remains valid, but $p_{\perp}$ is non-vanishing and large, so that now $\;R^{\mu}_{\;\;\mu}=-8\pi G T^{\mu}_{\;\;\mu}=16 \pi G(\rho-p_\perp)\;$. It follows that
\be
I_{EH,TL}  \;=\;
-4\pi \beta \int_{TL}~ dr  \; r^2   (\rho-p_\perp) + \mathcal{O}(\lambda)\; ,
\label{ITL}
\ee
where  $\;\beta=4\pi R\;$ is formally  a function of $r$; however, since the width of the layer is parametrically small, we can use that $\;r=R+{\cal O}(\lambda)\;$ and  ignore the $r$ dependence at leading order.

We now recall the conservation  equation~(\ref{conserv2})  and  also consider that,
in the layer, $\;p'_r \sim \dfrac{|p_r|}{\lambda}\gg |p_r|=\rho\;$, from which it can be deduced that $\;p_{\perp} \sim \dfrac{R}{\lambda}|p_r|\gg |p_r|=\rho\;$. Hence, we can ignore the contribution from $\rho$ in
Eq.~(\ref{ITL}) at leading order. Again invoking the conservation equation~(\ref{conserv2}) and   that the factors of $r$ appearing  in Eq.~(\ref{ITL}) can be approximated  as $R$'s, we then obtain
\be
I_{EH,TL}\;=\; 16{\pi^2} R\int\limits_{R(1-\lambda)}^{R(1+\lambda)}  dr
~\frac{R}{2}~ \partial_r(r^2 p_r)\;=\;
 8{\pi^2} R^2~ (r^2 p_r)\biggl|^{R(1+\lambda)}_{R(1-\lambda)}\;,
\label{ITL1}
\ee
up to order $\lambda/R$ corrections.

The value of $p_r$ at the upper limit in Eq.~(\ref{ITL1})  is its Schwarzschild value  $\;p_r =0\;$, while the value of  $p_r$ at the lower end is its frozen star (bulk) value
of $\;r^2 p_r=-\frac{1}{8\pi G}\;$, it follows that
\be
I_{EH,TL}\;=\; + \frac{\pi R^2}{G} + \mathcal{O}(\lambda)  \;.
\label{ITL2}
\ee

The remaining contribution is from the boundary term at infinity, which
gives the standard result for a Schwarzschild (exterior) spacetime,
\be
I_{EH,\infty}\;=\;-\frac{\pi R^2}{G}\;.
\label{fitty}
\ee

Summing up the results of this section from Eqs.~(\ref{lights,camera}),~(\ref{ITL2}) and~(\ref{fitty}), we find that the action (and thus the entropy \cite{U4Euclidean}) is
the same as it would be for a Schwarzschild BH of equal mass,
$\;I_{EH}= -\frac{\pi R^2}{G}+ \mathcal{O}(\lambda)\;$. However, for future reference,
what is important is that the
Einstein--Hilbert Euclidean action
for just the interior is vanishing,
\be
I_{EH,Interior}\;=\;I_{EH, Bulk}+I_{EH,TL}\;=\;0\;,
\label{IEH}
\ee
up to perturbative corrections.

\subsection{Born-Infeld action}\label{BI cont}

Let us now  evaluate the Born--Infeld contribution to the Euclidean action for the frozen star, starting with the bulk.  Recalling Eq.~(\ref{BILan}) for the
(Lorentzian) form of the action and taking into account that the constraint terms vanish on the solution, we find  that the Born--Infeld part of the Euclidean action  reduces to
\begin{equation}
	I_{BI}\;=\; \int d^3 x \oint dt_E\;\sqrt{g}\bigg\{E_i D^i+J_a A^a \bigg\}\;.
\end{equation}

In the bulk, there are no electric sources except at the center
of the star, so that the only other contribution comes from
\bea
\int d^3x \oint	dt_E \; \sqrt{g} E_i D^i \; &=&\;\frac{2q_{core}}{\alpha'}\int_{0}^{R-\lambda} r dr \cr
\;&=&\; \frac{q_{core} R^2}{\alpha'}+\mathcal{O}(\lambda)\;=\;\frac{\pi R^2}{G}+\mathcal{O}(\lambda)\;,
\eea
where we have used that $\;\oint dt_E=\beta(r)=4\pi r\;$, Eqs.~(\ref{DR},\ref{ER}) for the non-vanishing components of the fields and that,  in the last equality, $\;q_{core}=\frac{\pi \alpha'}{G}\;$.

As for the source at the center, this makes no contribution as can be seen  from
\begin{equation}
	\int d^3x \oint	dt_E \; \sqrt{g} J_a A^a \;=\; \int_{0}^{R-\lambda}  2r \frac{ q_{core}}{\alpha'}\delta(r)(r+C) dr\;=\;0\;,
	\label{STvanishes}
\end{equation}
where we  have used that $\;J_a= \delta_a^{\;\;0}J_0= q_{core}\delta(\vec{r})\;$ and $\;A^a=\delta^a_{\;\;0}A^0=\frac{r+C}{2\pi\alpha'} \;$ (here, $C$ is a constant of integration),
with the latter following  from
$\;E_r=\partial_r A^0\;$ and  $\;E_r=\frac{1}{2\pi\alpha'}\;$. The conclusion is
that
\be
{\cal I}_{BI,Bulk}\;=\;+ \frac{\pi R^2}{G}+\mathcal{O}(\lambda)\;.\label{act1}
\ee

Let us next consider the contribution from the transitional layer, starting with the electric-field  term,
\bea
\int	d^3x \oint dt_E\; \sqrt{g} E_i D^i &=& \int_{R-\lambda}^{R+\lambda} \frac{2rq_{core}}{\alpha'}dr\;\leq\;  \int_{R-\lambda}^{R+\lambda} 2r^3\frac{q_{core}}{\alpha' R^2} dr
\nonumber \\
&\leq &\frac{4q_{core}R}{\alpha'}\lambda +\mathcal{O}(\lambda^2)\;=\;\mathcal{O}(\lambda)\;.
\eea
In other words,  this  contribution  to the Euclidean action in the transitional layer is negligibly small.

We next consider the integral of the source term in the layer,
\begin{equation}
	\int d^3x  \oint dt_E \;\sqrt{g} J_a A^a \;=\;\int_{R-\lambda}^{R+\lambda}  16\pi^2 r^3 J_0 A^0 dr \;,
%\label{act2}
\end{equation}
for which the relation between the transverse pressure and the source in Eq.~(\ref{rel1}) allows the right-hand side to be rewritten as
\bea
-\int_{R-\lambda}^{R+\lambda}  16\pi^2 r^3 p_\perp dr &=&	-\int_{R-\lambda}^{R+\lambda} 16 \pi^2 r^3 \frac{1}{2r}
\partial_r(r^2 p_r)dr \nonumber \\
&=& -\int_{R-\lambda}^{R+\lambda} 8 \pi^2 R^2
\partial_r(r^2 p_r)dr  +  \mathcal{O}(\lambda) \label{act2}\\
&=&  8\pi^2 R^2 \;(r^2 p_r)_{R-\lambda} + \mathcal{O}(\lambda)
\;=\;-\frac{\pi R^2}{G}+\mathcal{O}(\lambda)\;,
\nonumber
\eea
where the conservation equation~(\ref{conserv2}) has been applied in the first line and the integrand has been rewritten as a total derivative in the second. As for the last line, that all matter densities vanish in the exterior
and  Eq.~(\ref{polypr}) for the frozen star pressure have been used
in the first and second equalities, respectively.
It then follows that
\be
I_{BI,TL}\;=\;-\frac{\pi R^2}{G} + \mathcal{O}(\lambda) \;.
\label{IBIS}	
\ee

As there are no external contributions to the Born--Infeld portion of the Euclidean
action, we can sum up  Eq.~(\ref{act1}) and  Eq.~(\ref{IBIS})
to arrive at
\be
I_{BI}\;=\;I_{BI, Bulk} + I_{BI, Source}\;=\;0 \;,
\label{IBI}	
\ee
up to perturbative-order corrections.

%%%%%%%%%%%%%%%%%%%%%%%%%%%%%%%%%%%%%%%%%%%%%%%%%

\subsection{The determinant prefactor}

We will now calculate the determinant prefactor
for the partition function~(\ref{gammy}) in the case of interest:
A collapsing matter shell with a Minkowski interior transitioning into a frozen star.

The standard procedure calls for the calculation of  these determinants by expanding the Euclidean action about the background solution of the equations of motion up to second order,
effectively converting the partition function into  a Gaussian integral.
Alternatively, for the frozen star (true vacuum), we will rather expand the equations of motion up to first order --- obtaining linear perturbation equations for the Einstein--Hilbert and Born--Infeld  actions --- and then show that  the perturbations are identically vanishing and therefore that $\;\text{det}'\;I''(\phi_{TV})=1\;$.  The same outcome applies trivially to the false vacuum, inasmuch as it is described by empty Minkowski space, so that
$\;\text{det}'\;I''(\phi_{FV})=1\;$.
The former result could have already  been anticipated from the ultrastability of the frozen star  geometry \cite{bookdill,popstar,fluctuations,frosty2}.

We will proceed  by first calculating the variations with respect to the metric only;  a very similar procedure to finding the equation for the propagation of gravitational waves on a curved background \cite{GW1,GW2}.  The variations with
respect to the Born--Infeld gauge fields will be subsequently considered.

\subsubsection{Metric perturbations}

The linearized Einstein equations are obtained by perturbing the frozen star background metric,
$\;g_{\mu\nu} \rightarrow g_{\mu\nu}+h_{\mu\nu}\;$, with $\;|h_{\mu\nu}|\ll 1\;$.
In order to simplify these equations and reduce the number of free parameters,
it is useful to implement the harmonic gauge,
	$\;g^{\mu\nu}\Gamma^{\rho}_{\mu\nu}[h]=0\;$,
\begin{equation}
	\frac{1}{2}g^{\mu\nu}g^{\rho\lambda}\left( \partial_\mu h_{\nu\lambda}+\partial_\nu h_{\lambda \mu}-\partial_{\lambda}h_{\mu\nu} \right)\;=\;0\;.
\end{equation}

Using that  $h_{\mu\nu}$ is symmetric  and choosing $\;h^{\nu}_{\;\nu}=0\;$, one
finds that the harmonic gauge
reduces to $\;\partial_{\mu}h^{\mu}_{\;
\lambda}=0\;$, which is also known as the Lorenz or transverse--traceless (TT) gauge.

\subsubsection{Transverse--traceless gauge for the frozen star geometry}

Given a general metric $g_{\mu\nu}$ for the background geometry  and stress  tensor $T_{\mu\nu}$
for the corresponding matter, it is not always possible to impose the TT gauge. Here, following \cite{Fanizza}, we will show that it is indeed possible to impose this gauge on
the frozen star geometry and stress tensor.
The TT gauge then allows us to choose the two non-vanishing components of the metric perturbation as  $h_{rr}$ and $h_{\theta r}$,
\begin{equation}
	h_{\mu\nu}\;=\; \begin{bmatrix}
		0 & 0 & 0&0\\
		0 & h_{rr} & h_{\theta r} &0&\\
		0 & h_{r \theta} & -h_{rr} &0\\
		0 &0&0&0
	\end{bmatrix}\;.
\label{hMuNu}
\end{equation}
Spherical symmetry of the perturbations will be restored eventually by showing that all the perturbations vanish.

 The linear metric perturbations $h_{\mu\nu}$ obey a dynamical evolution equation as follows \cite{Fanizza}:
\begin{equation}
	\begin{split}
		&\nabla^2 h_{\mu\nu}+2R_{\alpha\mu\beta\nu}h^{\alpha\beta}+R h_{\mu\nu}-g_{\mu\nu} h^{\alpha\beta}R_{\alpha\beta}-h^{\alpha}_{\mu}R_{\alpha\nu}-h^{\alpha}_{\nu}R_{\beta\mu}\\
		&-\nabla_\nu \nabla_\alpha h^{\alpha}_{\mu}-\nabla_\mu \nabla_\alpha h^{\alpha}_{\nu}+\nabla_\mu \nabla_\nu h-g_{\mu\nu}\left( \nabla^2 h -\nabla_\alpha \nabla_\beta h^{\alpha\beta}  \right)\;=\; -16\pi G \delta T_{\mu\nu}\;.
	\end{split}
\end{equation}

To be able to impose consistently the TT gauge, $\;\nabla_\nu h^{\mu\nu}=0\;$ and $\;h=g^{\mu\nu}h_{\mu\nu}=0\;$,  one finds that the following equation must be satisfied:
\begin{equation}
2R_{\mu\nu}h^{\mu\nu}\;=\;8\pi G g^{\mu\nu} \delta T_{\mu\nu}\;.
\label{condition1}
\end{equation}
Our goal is then to verify that condition~(\ref{condition1}) is satisfied for  the frozen star  geometry, thus justifying the use of Eq.~(\ref{hMuNu}). We will do so by showing that both sides of the equation vanish.

Let us begin with the right-hand side  of Eq.~(\ref{condition1}).
From Eq.~(\ref{BILan}), it follows that the contribution from the
main part of the Born--Infeld action
  to the stress tensor goes as
\be 
T^{BI}_{\mu\nu}\;=\; -g_{\mu\nu} E_i D^i + 2 E_i D_j \delta^i_{\;\;\mu} \delta^j_{\;\;\nu}\;,
\ee
so that
\be
\delta T^{BI}_{\mu\nu}\;=\; -h_{\mu\nu} E_i D^i\;.
\label{dtmunuED}
\ee
Additionally, from Eq.~(\ref{Tmunu}),  the Born--Infeld source term
makes a contribution of
\be
\delta T_{\mu\nu}^{source}\;=\; -h_{\mu\nu} J_0 A^0\;.
\label{duckyfuzz}
\ee
It is now clear that the right-hand side  of Eq.~(\ref{condition1}) is
vanishing because both of its contributions  are proportional  to the trace $\;h=0\;$.
 
We will next  show that the left-hand side    of Eq.~(\ref{condition1})
must also vanish because the relevant perturbations $h_{rr}$ and $h_{r\theta}$
both vanish.

\subsubsection{Perturbation equations and their solution}

By showing, in what follows, that all perturbations of the metric
are vanishing, we will be supporting  our claim that the linearized
Einstein equations  are identically zero, as well as validating our use
of the TT gauge.

With the TT gauge imposed, the linearized Einstein equations, without the contributions from the Born--Infeld sector, are found to be
\begin{equation}
	\Box h_{\mu\nu} + 2R_{\alpha\mu \beta\nu} h^{\alpha\beta}\;=\;0\;.
\end{equation}
In particular, the $r$--$r$ component of these equations is
\begin{equation}
	\Box h_{rr}\;=\;0\;,
\label{EOM1}
\end{equation}
as the second term vanishes because of the (anti-)symmetry properties of the Riemann tensor.
The $\theta$--$r$ component takes the form
\begin{equation}
		\Box h_{\theta r}+2R_{\theta rr\theta}h^{\theta r}\;=\;0\;.\label{EOM2}
\end{equation}

Let us now include  the contributions from the electric fields and their sources as described by Eqs.~(\ref{dtmunuED}) and~(\ref{duckyfuzz}), respectively.
The complete linearized equations
are then as follows:
\begin{equation}
	\Box h_{rr}-h_{rr} E_r D^r -J_0 A^0 h_{rr}\;=\;0
\label{Eom-rr}\;,
\end{equation}
\begin{equation}
	\Box h_{\theta r}+2 g^{rr}g^{\theta\theta} R_{\theta rr\theta}h_{\theta r} -h_{r\theta} E_r D^r
-J_0 A^0 h_{\theta r}=0
\label{Eom-rt}\;.
\end{equation}

We will choose our pair of boundary conditions for these equations such that  the innermost edge of the transitional layer (which is also the outer surface of the bulk of the frozen star)  is
left unperturbed. Then  $\;h_{\mu\nu}(t,r=R-\lambda)=0\;$ and $\;\partial_rh_{\mu\nu}(t,r=R-\lambda)=0\;$.
This leads to  $\;h_{r r}=h_{\theta r}=0\;$  on both sides of $r=R-\lambda$. Alternatively, we could have chosen the boundary conditions at the outer edge of the transitional layer, leading to the same results.

In the bulk, where the source term vanishes ({see Eq.~(\ref{STvanishes})),  Eq.~(\ref{Eom-rr}) reduces to
\be
-\frac{1}{\varepsilon^2} \partial_t^2 h_{rr}+{\varepsilon^2} \partial_r^2 h_{rr}+\Box_{\theta,\phi} h_{rr}-h_{rr} E_r D^r\;=\;0\;,
\label{Bhrreq}
\ee
where $\Box_{\theta,\phi}$ denotes the $\varepsilon^2$-independent angular part of the $\Box$ operator.
Multiplying Eq.~(\ref{Bhrreq}) by $\varepsilon^2$, one can see that it reduces to $\;\partial_t^2 h_{rr}=0\;$, whose solution with the prescribed boundary conditions is $\;h_{rr}=0\;$ to leading order in $\varepsilon^2$ and $\frac{\eta}{R}$.
Similarly, because $R_{\theta r r \theta}$ vanishes in the bulk,
it follows that $\;h_{\theta r}=0\;$ in the bulk at the same perturbative order.
We conclude that $\;h_{\mu\nu}=0\;$ throughout the bulk of the frozen star.

The situation  in the transitional layer is more complicated because
the non-angular components of the metric each become a power series in $\lambda$.
Nonetheless,
taking into account that the Riemann component $R_{\theta r r\theta}$
and derivative  $\partial_r$ both  scale as $1/\lambda$, whereas $\;\partial_t\sim 1/R\;$, one can obtain solvable equations. For instance,
after dropping terms that are obviously suppressed,
we find that the $r$--$r$ equation takes the form of
\be
-\partial^2_t h_{rr}+\partial^2_r h_{rr}+\partial_r h_{rr}+2h_{rr}\;=\;0\;,
\ee
where $\;r=R+{\cal O}(\lambda)\;$  has been used and the derivatives have been rescaled by powers of $R$ to render them dimensionless.
Notice that all the angular derivatives have been suppressed and, similarly, for the Born--Infeld fields and source terms.
The fields because  Eqs.~(\ref{DR}) and~(\ref{ER}) mean that
$E_rD^r$ scales as $\lambda^0$ and the source  because
$A^0J_0$ scales {\em only} as $1/\lambda$, {\em cf}, Eq.~(\ref{sourcey}).

The previous equation can readily be solved by separation of variables.  With the prescribed boundary conditions, the solution is simply $\;h_{rr}=0\;$.

Taking into account that $\;h_{rr}=0\;$ and imposing the above scaling relations,
one also obtains a tractable equation for $h_{\theta r}$,
\be
\partial_r h_{\theta r}+ h_{\theta r}\;=\;0\;,
\ee
whose solution, after imposing the boundary conditions is $\;h_{\theta r}=0\;$.

We conclude that $\;h_{\mu\nu}=0\;$ in the transitional layer of the frozen star. Combined with the result that $\;h_{\mu\nu}=0\;$ also in the bulk, it follows that all the gravitational perturbations vanish in the frozen star geometry, thus validating   the choice of the TT gauge.

More importantly, as the linearized equations of motion vanish identically, so too does the quadratic term in
the expansion of the frozen action. It then  follows
via $\;{\rm det}(A)=e^{{\rm Tr}\ln{A}}\;$ that the determinant of this
quadratic term should be unity.

\subsubsection{Gauge field perturbations}

Here, we show that perturbations in the Born--Infeld gauge fields make no contribution to the linearized equations of motion and, thus, no contribution to the determinant prefactor.

Varying $\;A^0 \rightarrow A^0 +\delta A^0\;$, we have $\;E_r=\partial_r (A^0 +\delta A^0)=\partial_r A^0 +\partial_r (\delta A^0)\;$.
So that, at linear order, the variation of the Born--Infeld  Lagrangian is the following:
\begin{equation}
	\frac{1}{\sqrt{g}}\frac{\delta\mathcal{L}_{BI}}{\delta A^0}\;=\; D^r \partial_r \delta A^0 + J_0 \delta A^0\;=\;\delta A^0\left(-\partial_r D^r+ J_0\right)\;=0\;,
\end{equation}
where the second equality required an integration by parts and the third equality is a consequence of the zeroth-order Gauss'-law constraint.
There is then no contribution to the linearized equations of motion from the fluctuations of $A^0$.

Similarly, the perturbation  $\;\widetilde{A}^0\rightarrow \widetilde{A}^0+\delta \widetilde{A}^0\;$ leads to
\begin{equation}
\frac{1}{\sqrt{g}}\frac{\delta\mathcal{L}_{BI}}{\delta \widetilde{A}^0}\;=\;g^{rr} E_r\; \partial_r \delta \widetilde{A}^0 \;,
\end{equation}
which via  the Euler--Lagrange equations yields
\begin{equation}
	\partial_r E^r\;=\;0\;,
\end{equation}
this being another zeroth-order result.
And so there is also no contribution from the fluctuations of $\widetilde{A}^0$.

As this set of  linearized equations vanishes identically, so too does the corresponding quadratic term in the expansion of the action; meaning
that the associated determinant will be unity.

%%%%%%%%%%%%%%%%%%%%%%%%%%%%%%%%%%%%%%%%%%%%%%%%%%%%%%%%%%%%%%%%%%%%%%%%%

%%%%%%%%%%%%%%%%%%%%%%%%%%%%%%%%%%%%%%%%%%%%%%%%%

\subsection{Transition probability}\label{TP}

Returning to Eq.~(\ref{gammy}), let us first consider the exponential factor.
Both spacetimes share an external Schwarzschild geometry, so that we need
only consider the respective interiors of the matter shell and the frozen star.
The action for empty Minkowski space is of course zero, $\;I_{FV}=0\;$. One might wonder about the contribution of the matter shell itself. But if we assume, for example,  a  thin shell of pressureless dust matter, then its Helmholtz free energy $F$ is
vanishing
and
likewise for its Euclidean action $I$ via the identification $\;I=-\beta F\;$ \cite{GH}.
 A different choice of realistic matter would lead
 to a contribution that, even if non-vanishing, can be expected to be on par with
 the  already neglected perturbative corrections. Meanwhile, using Eq.~(\ref{IEH}) and Eq.~(\ref{IBI}), one can see that the action for
 the frozen star interior also vanishes, $\;I_{TV}=0\;$, up to
 perturbative-order  corrections.
Hence, the exponential is  equal to unity up to perturbatively  small corrections.

Let us next consider the pre-exponential factor.
We have already shown that both determinants are unity.
 Because the tunneling event is triggered at a particular radius and thus at a particular time, as  determined by the  equation of motion for the infalling matter, there  can be no radial nor temporal zero modes  for either of the vacuua.  As  for the angular coordinates, their zero modes can either be fixed by a coordinate choice or, if not, these will be the same in both vacuua.  Any subtraction procedure due to the boundary at infinity must also cancel out because the two vacuua share the same external Schwarzschild geometry. As a result of these observations,
we conclude that the pre-exponential factor is equal to unity.

Since the partition function is unity,  it follows that
\begin{equation}
	\Gamma_{\rm{matter\;shell}\to{\rm frozen\;star}}\;=\;1\;,
\end{equation}
up to negligibly small corrections.
Meaning that, for  a collapsing shell of ordinary matter, the transition
into a frozen star of equal mass is an inevitable outcome!

Importantly, there is no longer any need to perform  Mathur's sum over microstates
\cite{mathur} to obtain our final result. This is because the bulk of the frozen star, whose entropy is $S_{BH}$ on its own, effectively performs the same summation when it is included in the Euclidean instanton.

%%%%%%%%%%%%%%%%%%%%%%%%%%%%%%%%%%%%%%%%%%%%%%%%%%

\section{Conclusion}

We have addressed what would be a key challenge for any  model of a BH mimicker
and explained how a frozen star can be formed from a shell of collapsing matter.
By adapting the  Euclidean-action method of Gibbons and Hawking and interpreting the action for an outer transitional layer as an instanton,
we have shown in a previous
article \cite{U4Euclidean} that
a collapsing body is likely to transition into a frozen star. Here, we improved on our previous results by showing that the  probability for a collapsing shell of matter to tunnel quantum mechanically into a frozen star is perturbatively close to unity and concluded that such a transition is therefore inevitable.

To obtain this result, we calculated the relevant Euclidean partition function, which  could then be identified with the  probability for the aforementioned transition to occur.  We showed that the difference in Euclidean actions between the true vacuum, the frozen star, and the false vacuum, the Minkowski interior of the shell, is a negligibly small number. And the determinant of small fluctuations around each of the respective solutions was be shown to be equal to unity. As a consequence of these findings, the partition function is itself equal to unity up to perturbatively small corrections.

As the gaps in the frozen star model continue to be filled, let us reemphasize that the ultimate goal is to make contact with observational physics; in particular,  through gravitational-wave observations. Relevant discussions along this line can be found in \cite{collision} for the polymer model and in \cite{frosty2} for the  frozen (actually, ``defrosted'') star.

\section*{Acknowledgments}
The research is supported by the German Research Foundation through a German-Israeli Project Cooperation (DIP) grant ``Holography and the Swampland'' and by VATAT (Israel planning and budgeting committee) grant for supporting theoretical high energy physics.
The research of AJMM received support from an NRF Evaluation and Rating Grant 119411. AJMM thanks Ben Gurion University for their hospitality on past visits. RB thanks the theory department at CERN and the department of theoretical physics at the University of Geneva for their hospitality.

%%%%%%%%%%%%%%%%%%%%%%%%%%%%%%%%%%%%%%%%%%%%%%%%%%%%%%%%

%%%%%%%%%%%%%%%%%%%%%%%%%%%%%%%%%%%%%%%%%%%%%%%%%%%%%%%%%%%%%%%%%%%%%%%%%%%%%%
%\bibliographystyle{unsrt}
%\bibliography{Ref.bib}

\begin{comment}

\end{comment}

\end{document}